\documentstyle[amstex,11pt]{article}

\begin{document}

\title{{\bf Charged Rotating Black Hole Formation from Thin Shell Collapse
in Three Dimensions}}
\author{Rodrigo Olea \medskip \\
{\small {\it Departamento de F\'{\i}sica, Pontificia Universidad Cat\'{o}%
lica de Chile, Casilla 306, Santiago 22, Chile.}} \and {\small {\it Centro
de Estudios Cient\'{\i}ficos (CECS), Casilla 1469, Valdivia, Chile}}.}
\date{}
\maketitle

\begin{abstract}
The thin shell collapse leading to the formation of charged rotating black
holes in three dimensions is analyzed in the light of a recently developed
Hamiltonian formalism for these systems. It is proposed to demand, as a way
to reconcile the properties of an infinitely extended solenoid in flat space
with a magnetic black hole in three dimensions, that the magnetic field
should vanish just outside the shell. The adoption of this boundary
condition results in an exterior solution with a magnetic field different
from zero at a finite distance from the shell. The interior solution is also
found and assigns another interpretation, in a different context, to the
magnetic solution previously obtained by Cl\'{e}ment and by Hirschmann and
Welch.
\end{abstract}

\section{Introduction}

General Relativity, as the geometrical theory describing the gravitational
interaction, possesses black hole solutions that are labeled by a small set
of parameters. This fact is a consequence of the No-Hair Theorem, that
states that the only relevant quantities are the mass, electric charge and
angular momentum. This assertion is valid if the matter stress tensor is
localized and that the fields are minimally coupled, assumptions that are at
any rate quite general. From a physical perspective, this idea is
interesting because just a few numbers are enough to completely classify the
solutions of the theory, even though they represent radically different
geometries.

A question that arises naturally is whether there exists a physical
mechanism that generates these solutions dynamically, for instance, through
gravitational collapse. One could also wonder if it is possible to reach the
whole space of solutions by means of such process, or if there are isolated
solutions (fundamental objects like extremal black holes), somehow
disconnected from the rest of the configurations. In this sense, the
validity of the Cosmic Censorship conjecture can also be tested in the
context of gravitational collapse (see, e.g., \cite{singh, harada}).

Among different types of gravitational collapse, thin shell collapse is an
interesting scenario to prove if naked singularities can be formed or not.
In this case, one studies the discontinuities between the inner and outer
geometries across a matter shell. The standard treatment for the collapse of
thin shells \cite{cruz-israel} has been successfully applied for more than
thirty years to analyze the dynamics of black hole creation and many other
related processes. Nevertheless, there may be some instances where this
method is hard to carry out in practice, for instance, when we are
interested in including rotation in these systems.

As an alternative approach to the Israel treatment of thin shell collapse, a
method based on the canonical formalism for gravity \cite{Regge:1974zd} has
been recently proposed \cite{COT1}. The direct integration of the
Hamiltonian constraints reproduces the standard shell dynamics for radial
collapse in a number of known cases. This work has also opened the
possibility of studying different rotating thin shell collapse cases in
three dimensions in a simpler manner. In this framework, the matching
conditions between the internal and external geometries are better
understood in Hamiltonian form, as discontinuities in the canonical momenta $%
\pi ^{ij}$.

This new Hamiltonian method was applied to obtain the dynamics of a
three-dimensional charged shell (the analogue of Reissner-Nordstrom collapse
in 4 dimensions \cite{Kuchar,Boulware}) and a rotating shell leading to the
formation of a BTZ black hole \cite{BTZ,BHTZ}. The general character of the
treatment in \cite{COT1} also proves the stability of the event horizon,
because a black hole cannot be converted into a naked singularity through
any thin shell collapse. Based on this method, one is able to conclude that
Cosmic Censorship always holds for any thin shell collapse. In particular,
this proof generalizes a result previously found for the electrically
charged case in 4 dimensions, that prevents the {\em overcharge} of a
Reissner-Norstrom black hole \cite{hubeny}.

The present paper studies the collapse of a charged rotating shell in three
spacetime dimensions, that creates a black hole with an external magnetic
field, as a further step in the application of this Hamiltonian approach.
Contrary to what one might think, this task turns out to be far from
straightforward, because it first requires to clarify several points
concerning the interior and exterior solutions, in order to select which of
them are compatible with the set-up.

Carrying out this procedure implies a nontrivial discussion about the
junction conditions for electromagnetism on the thin shell, as well. As we
shall see below, motivated by what happens with a solenoid in flat space (in
contradiction to the properties of charged rotating black holes in three
dimensions) it is proposed to demand that the magnetic field should vanish
just outside the shell throughout the collapse. The adoption of this
boundary condition results in a magnetic field which does not vanish in the
exterior (except as one becomes infinitesimally close to the shell) and
helps to reconcile both pictures (without and with gravity).

The exterior solution is the Mart\'{\i}nez-Teitelboim-Zanelli (MTZ) black
hole, obtained by a rotation {\em boost}\ in \cite{MTZ}, and can be obtained
from a family of solutions originally found in \cite{clement93} by certain
choice of the parameters. This class of exterior solutions possesses a
vanishing magnetic field at the horizon. The compatibility with the magnetic
boundary condition mentioned above is given by the fact that an external
observer will measure a vanishing magnetic field at the event horizon when
the shell crosses it. As the shell will appear infinitely red-shifted to
this distant observer, this boundary condition will ${\em freeze}$\ at the
horizon at the moment of the black hole creation.

On the other hand, the interior solution consistent with a
regularity condition at the center (that is also regular in the
limit of vanishing magnetic density) is a purely magnetic
spacetime (a vacuum solution
with a magnetic field, the analogue of Melvin solution in 4 dimensions \cite%
{melvin}). This geometry assigns another interpretation, in a different
context, to the magnetic solution previously obtained by Cl\'{e}ment \cite%
{clement93} and by Hirschmann and Welch \cite{HW}.

The work is organized as follows. Section 2 presents the Hamiltonian form of
the Einstein-Maxwell action and obtains the expressions for the reduced
action and field equations for rotational symmetry and time independence. As
the charged spinning shell divides the spacetime in two regions, it is
necessary to discuss suitable solutions for the interior and exterior
spacetimes, compatible with the presence of this source. This is done for
the interior solution in Section 3, where we study static magnetic
solutions. The analysis selects AdS spacetime with a magnetic field as the
only well-defined solution for the region enclosed by the shell. Section 4
briefly review some features of the charged rotating black hole discussed in
Ref.\cite{MTZ}, in the view to using it as the appropriate solution for the
external region. It also justifies our choice of boundary condition for a
vanishing magnetic field just outside the shell. The matching conditions for
electromagnetism across the shell are shown in Section 5, where the interior
magnetic flux is related to the parameters and dynamic variables of the
shell. Section 6 uses the method introduced in Ref.\cite{COT1} to study the
collapse of a charged spinning dust shell in three dimensions. Finally,
Section 7 summarizes the results presented here.

\section{Electrically Charged Rotating Solutions in (2+1) Dimensions}

In this section, we will briefly review the canonical formalism for gravity
coupled to electromagnetism in three spacetime dimensions.

The Einstein-Maxwell action for $(2+1)$ dimensions in presence of negative
cosmological constant is written as

\begin{equation}
I=\int d^{3}x\sqrt{-^{(3)}g}\left[ \frac{\left( R-2\Lambda \right) }{2\kappa
}-\frac{1}{4}F_{\mu \nu }F^{\mu \nu }\right] .  \label{I(2+1)QJ}
\end{equation}
Here, the cosmological constant is $\Lambda =-\frac{1}{l^{2}}$ in terms of
the AdS length. We will adopt the units of Ref.\cite{MTZ}, i.e., the
gravitational constant $\kappa =8\pi G$ is set equal to $\frac{1}{2}$.

The line element for a timelike ADM foliation \cite{ADM}, in terms of the
lapse $N^{\perp }$ and shift $N^{i}$ Lagrange multipliers, and the spatial
two-dimensional metric $g_{ij}$ is written as

\begin{equation}
ds^{2}=-(N^{_{\bot }})^{2}dt^{2}+g_{ij}(N^{i}dt+dx^{i})(N^{j}dt+dx^{j}).
\label{ds}
\end{equation}%
Then, the spacetime is described by an infinite series of constant-time
slices $\Sigma _{t}$ generated by the time-like normal vector $n_{\mu
}=(-N^{_{\bot }},0,0)$.

We take as canonical variables for the gravitational part the metric $g_{ij}$
and its conjugate momenta $\pi ^{ij}$ and, for the electromagnetic sector,
the vector potential $A_{i}$ and its conjugate momentum ${\cal E}^{i}$ \cite%
{Regge:1974zd}. This choice leads to the Hamiltonian form of the action (\ref%
{I(2+1)QJ})

\begin{equation}
I=\int dtd^{2}x\left( \pi ^{ij}\dot{g}_{ij}+{\cal E}^{i}\overset{.}{A_{i}}%
-N^{_{\bot }}{\cal H}_{\bot }-N^{i}{\cal H}_{i}-A_{0}G\right)
\end{equation}%
where the components of the Hamiltonian and the gauge transformations
generator $G$ are given by

\begin{eqnarray}
{\cal H}_{\bot } &=&\frac{2\kappa }{\sqrt{g}}\left( \pi ^{ij}\pi
_{ij}-\left( \pi _{i}^{i}\right) ^{2}\right) -\frac{\sqrt{g}}{2\kappa }%
\left( ^{\left( 2\right) }\!R(g)+\frac{1}{l^{2}}\right)  \label{hper} \\
&&+\frac{1}{2\sqrt{g}}\left( g_{ij}{\cal E}^{i}{\cal E}^{j}+{\cal B}%
^{2}\right) +\sqrt{g}T_{\bot \bot }  \nonumber \\
{\cal H}_{i} &=&-2\pi _{i\mid j}^{j}+\epsilon _{ij}{\cal E}^{j}{\cal B}+%
\sqrt{g}T_{\bot i}  \label{hi} \\
G &=&-\left( {\cal E}^{i},_{i}\right) .
\end{eqnarray}

Here we have defined the magnetic density as

\begin{equation}
F_{ij}=\epsilon _{ij}{\cal B}.
\end{equation}

For stationary, circularly symmetric solutions of the form

\begin{equation}
ds^{2}=-N^{2}(r)F^{2}(r)dt^{2}+F^{-2}(r)dr^{2}+r^{2}(d\varphi +N^{\varphi
}(r)dt)^{2}  \label{ansatzqj}
\end{equation}%
with $0\leq r<\infty $ and $0\leq \varphi <2\pi $, there is only one
nonvanishing component of the gravitational momentum, that is,

\begin{equation}
\pi _{\varphi }^{\;r}=p(r).  \label{piphir}
\end{equation}%
In the electromagnetic sector, due to rotational symmetry, the magnetic
density becomes

\begin{equation}
{\cal B}=\partial _{r}A_{\varphi }  \label{magndens}
\end{equation}%
that is related to the magnetic field referred to an orthonormal basis $B$ by

\begin{equation}
B=g^{-1/2}{\cal B}  \label{Bmag}
\end{equation}%
The Gauss law is written as

\begin{equation}
{\cal E}^{r}=Q  \label{electricfield}
\end{equation}
with $Q$ the total electric charge.

The reduced action, coming from the direct substitution of the metric ${\em %
Ansatz}$ (\ref{ansatzqj}) and the fields (\ref{piphir})-(\ref{electricfield}%
) in the Hamiltonian form of the action (\ref{I(2+1)QJ}) can be expressed as

\begin{equation}
I=-2\pi \Delta t\int \left( N{\cal H}+N^{\varphi }{\cal H}_{\varphi }\right)
dr,  \label{reducedI}
\end{equation}%
after a redefinition of the generator of normal deformations ${\cal H}=F%
{\cal H}_{\bot }$ and due to the fact that the fields depend only on the
radial coordinate. Altogether, the angular component of the Hamiltonian
takes the form

\begin{equation}
{\cal H}_{\varphi }=-2p^{\prime }-Q{\cal B}
\end{equation}%
where the prime denotes a derivative respect the radial coordinate.

The variation of the reduced action (\ref{reducedI}) with respect the
relevant fields, i.e., $N$, $N^{\varphi }$, $F^{2}$, $p$ and $A_{\varphi }$
leads to the set of equations

\begin{eqnarray}
(F^{2})^{\prime }-2r+\frac{2p^{2}}{r^{3}}+\frac{Q^{2}}{2r}+\frac{F^{2}{\cal B%
}^{2}}{2r} &=&0  \label{Fprime} \\
2p^{\prime }+Q{\cal B} &=&0  \label{varNphi} \\
N^{\prime }-\frac{N{\cal B}^{2}}{2r} &=&0  \label{varFsquare} \\
(N^{\varphi })^{\prime }+\frac{2Np}{r^{3}} &=&0  \label{varp} \\
\left( \frac{NF^{2}{\cal B}}{r}-N^{\varphi }Q\right) ^{\prime } &=&0.
\label{varAphi}
\end{eqnarray}%
For the problem of gravitational collapse of a thin shell that we shall
discuss below, we will simply consider the spacetime divided in an interior
and exterior regions. Thus, the above equations hold at either side of the
shell with all the variables carrying the respective subscript.

We also include in the r.h.s. of Eqs.(\ref{hper},\ref{hi}) a matter
energy-momentum tensor for the shell, as a source for the discontinuities of
the spacetime geometry across it. In this way, one is able to study the
dynamic evolution of the thin shell in Hamiltonian form. In Ref. \cite{COT1}%
, this procedure was carried out for a perfect fluid, with a stress tensor
that is not derived from the action (\ref{I(2+1)QJ})%
\begin{equation}
T_{\mu \nu }=\left[ \sigma u_{\mu }u_{\nu }-\tau (h_{\mu \nu }+u_{\mu
}u_{\nu })\right] \delta (X)  \label{Tmunu}
\end{equation}%
where $u^{\mu }$ is the shell velocity, $\sigma $ the energy density and $%
\tau $ the tension. The delta function gives a matter distribution localized
at the boundary of the hypersurface defined by the shell evolution. In
particular, in three-dimensional spacetimes one is able to obtain the
equations governing the collapse of a rotating thin shell.

Finally, we know that in general the Hamiltonian action principle for
gravity is not well-defined without boundary terms. In the context of thin
shell collapse we will assume that the action is supplemented by the
appropriate boundary terms such that the action is stationary {\em on-shell }%
\cite{Regge:1974zd}, and that permits the computation of the correct
conserved charges in the asymptotic region. Thus, the mass, angular momentum
and electric charge of the external solution come out naturally as
quantities measured at spatial infinity. In turn, the same parameters for
the internal solution, and the relation to the exterior charges, will be
identified only when the shell dynamics is solved.

\section{Static Magnetic Solutions in $3$ Dimensions}

For simplicity, one might set the internal solution for the collapse process
as a static BTZ black hole. However, if the surrounding shell is rotating,
it generates an internal magnetic flux whose equivalent gravitational energy
regularizes the inner space, erasing the formerly present event horizon. The
geometry of this spacetime was first found by Cl\'{e}ment in \cite{clement93}
and its properties were studied in extenso by Hirschmann and Welch in \cite%
{HW}. In the latter reference, the authors carry out a detailed analysis to
make manifest the absence of black holes for any value of the magnetic
density and mass $M$. Here, we briefly review this discussion, to finally
conclude that for $M>0$, the solution is in general ill-defined in the limit
of a nonrotating shell.

The line element for this solution takes the explicit form\footnote{%
Henceforth, we set the AdS radius equal to unity.}

\begin{equation}
ds^{2}=-\left( \rho ^{2}-M\right) dt^{2}+\frac{\rho ^{2}}{\left( \rho
^{2}-M\right) }\frac{d\rho ^{2}}{\left( \rho ^{2}+\frac{\Theta ^{2}}{4}\ln
\left\vert \rho ^{2}-M\right\vert \right) }+\left( \rho ^{2}+\frac{\Theta
^{2}}{4}\ln \left\vert \rho ^{2}-M\right\vert \right) d\phi ^{2}
\label{hwelch}
\end{equation}%
where $M$ is the original BTZ black hole mass and $\Theta $ is an
integration constant (that is interpreted as a magnetic charge in \cite{HW}%
.). From the Eq.(\ref{Bmag}), we obtain the form of the inner magnetic field
as

\begin{equation}
B_{-}(\rho )=\frac{\Theta }{\sqrt{\rho ^{2}-M}}.  \label{bminus}
\end{equation}

In the case of gravitational collapse we are interested in, $\Theta $
represents the magnetic flux generated by the shell and it will depend on
the shell parameters and dynamic variables. As the external magnetic field
is given (we assume the outer mass, angular momentum and charged as known)
the internal magnetic field is determined using junction conditions for
electromagnetism across the shell.

The spacetime regularization mentioned above remains valid even for a
however small amount of magnetic flux, {\em i.e.}, the slightest addition of
angular momentum to the shell wipes the horizon out.

This can be understood by changing the form of the metric with the use of a
new coordinate $x^{2}=\rho ^{2}-\bar{\rho}^{2}$, that covers completely the
spacetime. The radius $\bar{\rho}$ is given by the position where $g_{\phi
\phi }$ vanishes, i.e., the one that solves the transcendental equation

\begin{equation}
\bar{\rho}^{2}-M=e^{-4\bar{\rho}^{2}/\Theta ^{2}}.  \label{rhobar}
\end{equation}%
With this coordinate choice and the definition $\delta ^{2}=\bar{\rho}^{2}-M$%
, the metric near $x=0$ is seen as a conical spacetime

\begin{equation}
ds^{2}=-\left( x^{2}+\delta ^{2}\right) dt^{2}+\frac{dx^{2}}{\left(
x^{2}+\delta ^{2}\right) \left( 1+\frac{\Theta ^{2}}{4\delta ^{2}}\right) }%
+x^{2}\left( 1+\frac{\Theta ^{2}}{4\delta ^{2}}\right) d\phi ^{2}.
\label{hwelchx0}
\end{equation}%
In order to avoid a conical singularity, the angular coordinate needs to be
identified in a period

\begin{equation}
T_{\phi }=2\pi \frac{e^{2\overline{\rho }^{2}/\Theta ^{2}}}{1+\frac{\Theta
^{2}}{4}e^{4\overline{\rho }^{2}/\Theta ^{2}}}.  \label{period}
\end{equation}

Nevertheless, we have to keep in mind that we are considering this solution
in the context of a rotating shell collapse and the above picture must be
consistent also when the shell is falling radially. In spite of the form of
the line element (\ref{hwelch}), for any solution with $M>0$ the limit $%
\Theta \rightarrow 0$ does not recover the spinless BTZ metric \cite%
{BTZ,BHTZ} because the angular period (\ref{period}) approaches zero. On the
contrary, for this $M<0$ the value of $\overline{\rho }$ from Eq.(\ref%
{rhobar}) can be expanded as%
\begin{equation}
\overline{\rho }^{2}=-\frac{\Theta ^{2}}{4}\ln (-M)+O(\Theta ^{4})
\end{equation}%
that has a regular limit when $\Theta $ goes to zero, and the period (\ref%
{period}) goes to a constant. Therefore, demanding that the solution is
compatible with the nonrotating case makes us consider only solutions with $%
M<0$. This is also in agreement with the work by Dias and Lemos \cite%
{DiasLemos}, where they have proved that the mass of the three-dimensional
static magnetic solution is necessarily negative.

However, it has been pointed out in \cite{cataldoetal} that the mass of the
magnetic solution can always be rescaled to $M=-1$ using a set of
appropriate transformations and that this spacetime is actually a
one-parameter solution.

The above argument implies that the magnetic AdS spacetime is the only
internal solution with physical relevance for this set-up of thin shell
collapse. This solution may be considered the analogue of Melvin spacetime
\cite{melvin} for 3 dimensions, as it represents a vacuum configuration with
a magnetic field filling the space up. This process of regularization of the
magnetic solution in three dimensions had been originally found in \cite%
{clement93} for some particular values of the parameters.

Thus, the previous reasoning assigns another interpretation to the solution
of Cl\'{e}ment-Hirschmann-Welch, where now the parameter $\Theta $ is
understood as a magnetic flux generated by a surrounding solenoid.

For $M=-1$, the radius $\bar{\rho}$ takes the value $\bar{\rho}=0$.
Therefore, a simple redefinition in the angle $\varphi =\left( 1+\frac{%
\Theta ^{2}}{4}\right) \phi $ permits us to absorb the angular deficit in
the metric (\ref{hwelchx0}) in the limit $x\rightarrow 0$, avoiding the
presence of a conical singularity. The line element for this solution takes
the form

\begin{equation}
ds^{2}=-\left( \rho ^{2}+1\right) dt^{2}+\frac{\rho ^{2}}{\left( \rho
^{2}+1\right) }\frac{d\rho ^{2}}{\left( \rho ^{2}+\frac{\Theta ^{2}}{4}\ln
\left( \rho ^{2}+1\right) \right) }+\frac{\left( \rho ^{2}+\frac{\Theta ^{2}%
}{4}\ln \left( \rho ^{2}+1\right) \right) }{\left( 1+\frac{\Theta ^{2}}{4}%
\right) ^{2}}d\varphi ^{2}.  \label{COmagnetic}
\end{equation}

In sum, the magnetic solution (\ref{COmagnetic}) derived from the
discussion in this section is determined by the conditions of
circular symmetry, the regularity at the center and the regularity
of the nonrotating shell limit. This last criterion could be of
particular relevance in a system that incorporates dissipative
forces that could eventually bring the rotation to a halt.

The analysis of the discontinuities of this geometry with the exterior
solution (MTZ black hole) will consider two classes of junction conditions.
The first class deals with the jump in the metric tensor and extrinsic
curvature across the thin shell, as an application of a recent work that
treats rotating shell collapse in three dimensions in Hamiltonian form \cite%
{COT1}.

Expressing the metric (\ref{COmagnetic}) in the form of Eq.(\ref{ansatzqj})
we study the matching conditions for the functions $N(r)$ and $N^{\varphi
}(r)$ across the shell

\begin{equation}
ds^{2}=-N_{-}^{2}(r)F_{-}^{2}(r)dt^{2}+\frac{dr^{2}}{F_{-}^{2}(r)}%
+r^{2}d\varphi ^{2}.
\end{equation}%
In terms of the new radial coordinate $r$

\begin{equation}
r^{2}=\frac{(\rho ^{2}+\frac{\Theta ^{2}}{4}\ln (\rho ^{2}+1))}{(1+\frac{%
\Theta ^{2}}{4})^{2}},  \label{r_rho}
\end{equation}%
the functions $N_{-}$ and $F_{-}$ in the metric are given by

\begin{equation}
N_{-}^{2}=\frac{\left( \rho ^{2}+1\right) ^{2}\left( 1+\frac{\Theta ^{2}}{4}%
\right) ^{2}}{\left[ \rho ^{2}+\left( 1+\frac{\Theta ^{2}}{4}\right) \right]
^{2}}
\end{equation}%
and%
\begin{equation}
F_{-}^{2}=\frac{\left[ \rho ^{2}+\left( 1+\frac{\Theta ^{2}}{4}\right) %
\right] ^{2}}{\left( \rho ^{2}+1\right) \left( 1+\frac{\Theta ^{2}}{4}%
\right) ^{2}}.  \label{Fminus}
\end{equation}

In the same way as we discussed in Ref.\cite{COT1}, we can always
reparametrize the time such that $N_{-}(r=R)=1$, because $N_{-}$ is defined
up to a constant. With this choice we ensure the continuity of $N(r)$ across
the thin shell, and we measure the jump in $F^{2}(r)$. Then, the internal
magnetic field (\ref{bminus}) takes the particular form on the shell position%
\begin{equation}
B_{-}(R)=\frac{\Theta }{F_{-}(R)}.  \label{Bminuspart}
\end{equation}

We also demand the angular shift $N^{\varphi }(r)$ to be continuous at $r=R$%
. As the interior spacetime is static, a consistent choice for the
shift in the exterior region is the one that satisfies
$N_{+}^{\varphi }(r=R)=0$ at all times. This corresponds to a
reparametrization in the angular variable, that differs from the
standard value $N^{\varphi }(r=\infty )=0$ that the leaves
the infinity nonrotating. A similar choice has been considered in \cite%
{HHT-R} to compute the conserved quantities of Kerr-AdS black holes referred
to a frame rotating at infinity.

The second class of junction conditions considers the discontinuities in the
Maxwell field strength. As we will discuss in the next section, the
compatibility of the magnetic properties between the static interior and
rotating exterior solutions will motivate the introduction of an additional
boundary condition for the magnetic field.

\section{External Solution}

At first sight, one might be puzzled because of the different qualitative
behavior of a magnetic field in flat and curved spacetimes.

Let us consider an infinitely extended charged cylinder in four dimensions.
In flat space, if the cylinder is rotating, it generates an internal
magnetic field. This magnetic field vanishes just outside this solenoid, and
keeps that value throughout the exterior region. This property is not
modified if the cylinder is also collapsing as it rotates: the inner
magnetic field just changes with time as the radius decreases.

Contrary to the above situation, when electromagnetism is coupled to gravity
in 3 dimensions, a charged rotating black hole solution possesses an
external magnetic field $B_{+}(r)$. For a black hole exterior metric of the
form (\ref{ansatzqj}), the magnetic field (\ref{Bmag}) is%
\begin{equation}
B_{+}={\cal B}\frac{F_{+}(r)}{r}
\end{equation}%
and therefore, $B_{+}$ in general will be different from zero everywhere but
at the event horizon \cite{MTZ}.

Roughly speaking, that means that turning on the gravitational field in a
flat space with a charged rotating ring modifies the external magnetic field
such that is not longer zero outside.

As a way to reconcile both pictures, we demand the magnetic field
to vanish just outside the rotating shell during the entire
collapse. This choice is compatible with a nonvanishing $B_{+}$ at
a finite distance from the shell. As this boundary condition will
hold throughout the collapse, in particular when this ring crosses
its own event horizon $r_{+}$, the process will finally leave as a
remnant a magnetic field which does not vanish in the exterior
region, only at $r=r_{+}$. A similar choice of a boundary
condition occurs in the rotating case studied in Ref.\cite{COT1},
where the choice of the angular shift as $N_{+}^{\varphi }(r=R)=0$
produces an exterior solution with angular momentum different from
zero.

In the present case, the result is an exterior black hole, which
features a magnetic field that satisfies this boundary condition.
This solution is precisely the one obtained from a charged
($\tilde{M}\neq 0,$ $\tilde{Q}\neq 0$) static solution
\begin{equation}
f^{2}(\tilde{r})=\tilde{r}^{2}-\tilde{M}-\frac{1}{4}\tilde{Q}^{2}\ln \tilde{r%
}^{2}
\end{equation}%
by a Lorentz boost in the $t-\varphi $ plane, an illegitimate coordinate
transformation that changes the physical parameters \cite{MTZ}. This
transformation is characterized by a boost parameter $\omega $, that
corresponds to an angular velocity. A similar procedure had been used
previously to obtain the rotating BTZ black hole metric from the static one
\cite{Clement96}.

By means of a redefinition in the radial variable

\begin{equation}
r^{2}=\tilde{r}^{2}+\frac{\omega ^{2}}{\left( 1-\omega ^{2}\right) }\left(
\tilde{M}+\frac{\tilde{Q}^{2}}{4}\ln \tilde{r}^{2}\right)  \label{r_rtilde}
\end{equation}%
we obtain the expressions for the functions in the metric

\begin{eqnarray}
F_{+}^{2} &=&\frac{\left( \tilde{r}^{2}+\frac{\omega ^{2}\tilde{Q}^{2}}{%
4\left( 1-\omega ^{2}\right) }\right) ^{2}}{r^{2}\tilde{r}^{2}}\left( \tilde{%
r}^{2}-\tilde{M}-\frac{1}{4}\tilde{Q}^{2}\ln r^{2}\right)  \label{Fplus} \\
N_{+} &=&\frac{\tilde{r}^{2}}{\tilde{r}^{2}+\frac{\omega ^{2}\tilde{Q}^{2}}{%
4\left( 1-\omega ^{2}\right) }} \\
N_{+}^{\varphi } &=&-\omega ^{2}\frac{\tilde{M}+\frac{\tilde{Q}^{2}}{4}\ln
\tilde{r}^{2}}{\left( 1-\omega ^{2}\right) r^{2}}
\end{eqnarray}%
and for the magnetic density%
\begin{equation}
{\cal B}_{+}=\frac{\tilde{Q}\omega r}{\sqrt{1-\omega ^{2}}\left( \tilde{r}%
^{2}+\frac{\omega ^{2}\tilde{Q}^{2}}{4\left( 1-\omega ^{2}\right) }\right) }.
\end{equation}

From the asymptotic from of the above equations, one is able to identify the
mass, angular momentum and electric charge of the solution generated by the
boost%
\begin{eqnarray}
M &=&\frac{1}{1-\omega ^{2}}\left( \tilde{M}\left( 1+\omega ^{2}\right) -%
\frac{\omega ^{2}\tilde{Q}^{2}}{2}\right) ,  \label{Mfromtildes} \\
J &=&\frac{2\omega }{1-\omega ^{2}}\left( \tilde{M}-\frac{\tilde{Q}^{2}}{4}%
\right) ,  \label{Jfromtildes} \\
Q &=&\frac{\tilde{Q}}{\sqrt{1-\omega ^{2}}},  \label{Qfromtildes}
\end{eqnarray}%
where the parameter $\omega $ is the root of the cubic equation%
\begin{equation}
\frac{Q^{2}}{4}\omega ^{3}-\frac{J\omega ^{2}}{2}+\omega \left( M-\frac{Q^{2}%
}{4}\right) -\frac{J}{2}=0.  \label{omegacubic}
\end{equation}

\section{Junction Conditions for Electromagnetism}

In this section we will analyze the junction conditions for electromagnetism
across the thin shell. In order to do that, we will consider a decomposition
of the Maxwell equations along the normal and tangent directions to the
shell to study the discontinuity in the electromagnetic field strength.

The spacetime is divided in two regions, described by the sets $\left\{
x_{\pm }^{\mu }=t_{\pm },r_{\pm },\varphi \right\} $ in Schwarzschild-like
coordinates. The hypersurface $\Sigma _{\xi }$ (normal to a unit vector $\xi
^{\mu }$) is generated by the collapsing shell is the boundary and separates
both spaces, where we define the induced metric $h_{ab}$, the intrinsic
coordinates set $\left\{ \sigma ^{a},a=0,2\right\} $ and the projector
between these two bases $e_{a}^{\mu }=\frac{\partial x^{\mu }}{\partial
\sigma ^{a}}$.

In the Ref.\cite{COT1} it was shown that a convenient way to study the
discontinuities in the metric in the rotating case is adapting the intrinsic
geometry to a frame falling radially (nonrotating) with the shell. An
observer in this frame measures a proper time $\lambda $, that differs from
the time coordinate $\tau $ of an observer rotating with the shell
\begin{equation}
d\lambda =\gamma d\tau
\end{equation}%
where $\gamma $ corresponds to the relativistic factor associated to the
angular velocity $\Omega =d\varphi /d\lambda $

\begin{equation}
\gamma =\frac{1}{\sqrt{1-R^{2}\Omega ^{2}}}.
\end{equation}%
The commoving observer sees an intrinsic geometry given by the shell line
element
\begin{equation}
ds^{2}=h_{ab}d\sigma ^{a}d\sigma ^{b}=-d\lambda ^{2}+R^{2}(\lambda )d\varphi
^{2}
\end{equation}%
and the three velocity in this frame takes the explicit form

\begin{equation}
u^{\mu }=\gamma \left( \frac{\alpha }{F^{2}},\dot{R},\Omega \right)
\end{equation}%
where the dot denotes a derivative respect the time $\lambda $, the function
$\alpha $ is defined as
\begin{equation}
\alpha =\sqrt{F^{2}+\dot{R}^{2}}  \label{alphadef}
\end{equation}%
So, this description considers a nonrotating orthonormal system $\{\lambda
,X\}$, where $\lambda $ is the tangential coordinate that runs along the
nonrotating components of $u^{\mu }$ and $X$ goes along the normal $\xi
^{\mu }$. Therefore, any displacement in the plane $\{t,r\}$ of
Schwarzschild-like coordinates can be equivalently expressed as

\begin{eqnarray}
dt &=&u^{t}d\lambda +\xi ^{t}dX  \label{dt} \\
dr &=&u^{r}d\lambda +\xi ^{r}dX  \label{dr}
\end{eqnarray}%
expressions that will be useful to project the components of the strength
tensor $F_{\mu \nu }$ on $\Sigma _{\xi }$.

For the components along this hypersurface, the discontinuity can be
expressed as

\begin{equation}
\left[ F_{ab}\right] =\left. F_{ab}\right| _{+}-\left. F_{ab}\right| _{-}=0
\label{disFab}
\end{equation}
and for the remaining ones

\begin{equation}
\left[ F_{a\xi }\right] =-j_{a}  \label{disFaxi}
\end{equation}%
where we have defined the projections $F_{ab}=e_{a}^{\mu }e_{b}^{\nu }F_{\mu
\nu }$ and $F_{a\xi }=e_{a}^{\mu }\xi ^{\nu }F_{\mu \nu }$, respectively.
The r.h.s. of eq.(\ref{disFaxi}) represents the electric current $j_{a}$
generated by the shell rotation.

The current can be written in terms of the three-velocity
\begin{equation}
j_{a}=\eta e_{a\mu }u^{\mu }  \label{ja}
\end{equation}%
where the constant $\eta $ is the charge density of the ring, such that

\begin{equation}
q=2\pi R\eta \gamma
\end{equation}%
is the total charge in the commoving frame.

Substituting the above formulas in the only nonvanishing component of (\ref%
{disFab}), produces the expression

\begin{equation}
\left[ F_{02}\right] =0=\left[ \frac{\alpha }{F^{2}}F_{t\varphi }\right] +%
\left[ \dot{R}F_{r\varphi }\right] .  \label{Fab02}
\end{equation}%
The jump in the magnetic field (\ref{Bmag}) across the shell is related to
the induced $\varphi $-component of the electric field

\begin{equation}
\left[ \dot{R}\frac{{\cal B}F}{R}\right] =-\left[ \frac{\alpha }{F^{2}}%
F_{t\varphi }\right]  \label{BEphi}
\end{equation}%
that, in principle, it is not necessarily zero in three dimensions. In fact,
Maxwell equations tell us that the azimuthal electric field can be different
from zero just outside the shell (wherever the magnetic field vanishes \cite%
{cataldo})

On the other hand, the $a=0$ component of eq.(\ref{disFaxi}) leads to

\begin{equation}
\left[ F_{tr}\right] =\frac{q}{2\pi R}.
\end{equation}
Due to the absence of electric charge in the inner space, the last equation
is just the Gauss law for the exterior electric field

\begin{equation}
{\cal E}_{+}^{r}=\frac{q}{2\pi }.  \label{Gauss}
\end{equation}
The other relevant equation corresponds to the angular component of (\ref%
{disFaxi})

\begin{equation}
\left[ F_{2\xi }\right] =-\left[ \frac{\dot{R}}{F^{2}}F_{t\varphi }\right] -%
\left[ \alpha \frac{{\cal B}F}{R}\right] =-R^{2}\eta \Omega \gamma
\end{equation}%
that, using (\ref{BEphi}), can be write finally as%
\begin{equation}
\left[ \frac{F^{2}}{\alpha }B\right] =\frac{\Omega qR}{2\pi }.  \label{Bjump}
\end{equation}

The above formula can be regarded as the analogue of Ampere's law in $(2+1)$
dimensions. Because the magnetic field is a scalar, we can imagine it
pointing along the time direction. Tracing an infinitesimal `rectangular'
contour ${\cal C}$ that crosses twice --forth and back-- the hypersurface $%
\Sigma $, the bottom and top lines (contained in constant-time slices at $%
t=t_{0}$ and $t=t_{0}+dt$) do not contribute to the integral of the magnetic
field along the path. The integration along the time gives the magnetic
field $B$ multiplied by $dt^{\perp }=dtN^{\perp }n$, the normal timelike
distance that separates the slices (see, e.g., \cite{Regge:1974zd}), with
the corresponding subscript at each side of the hypersurface $\Sigma _{\xi }$%
\begin{equation}
\oint\limits_{{\cal C}}B.dt=\left( N_{+}^{\perp }\right)
^{2}B_{+}dt_{+}-\left( N_{-}^{\perp }\right) ^{2}B_{-}dt_{-}=I
\label{closeint}
\end{equation}%
where $I$ is the total current enclosed by the contour ${\cal C}$.

An observer falling radially with the shell measures distances $\left\{
\lambda ,X\right\} $ along the unit velocity $u^{\mu }$ and normal $\xi
^{\mu }$, that are related to the coordinates $\{t,r\}$ of an external
observer by Eqs.(\ref{dt}) and (\ref{dr}). Therefore, an infinitesimal time
displacement $dt$ at constant $r$ ($dr=0$) is expressed as $dt=\frac{%
d\lambda }{\alpha }$. Recalling that $N^{\perp }=NF$ then, infinitesimally
close to the shell $N=1$ and therefore

\begin{equation}
\left( \frac{F_{+}^{2}}{\alpha _{+}}B_{+}-\frac{F_{-}^{2}}{\alpha _{-}}%
B_{-}\right) d\lambda =I  \label{ampere}
\end{equation}%
The total current in the l.h.s. of Eqs.(\ref{closeint},\ref{ampere}) is
given by the product between the azimuthal component of the current density (%
\ref{ja}) and the proper time distance $d\lambda $ measured along $\Sigma
_{\xi }$,%
\begin{equation}
I=\gamma R^{2}\eta \Omega d\lambda
\end{equation}%
that exactly reproduces the result of Eq.(\ref{Bjump}).

Finally, the boundary condition introduced above requires the external
magnetic field to vanish just outside the shell, and therefore

\begin{equation}
\frac{F_{-}^{2}}{\alpha _{-}}B_{-}=-\frac{\Omega qR}{2\pi }.  \label{Bminusq}
\end{equation}

\section{Collapse Dynamics for a Dust Thin Shell}

In a recent paper, a Hamiltonian approach to treat the gravitational
collapse of rotating thin shells in three dimensions has been introduced
\cite{COT1}.

This procedure is probably equivalent to the Israel junction conditions for
rotating shells. However, what remains unclear in the Israel Lagrangian
method is how to define appropriate rotating frames for the inner and outer
spaces and for the shell itself to match the different geometries.

In the Hamiltonian framework, the continuity at the shell position $r=R$ of
the (redefined) lapse function $N$ and the angular shift $N^{\varphi }$ in
Eq.(\ref{ansatzqj}) allows us to compute the in the canonical momenta $\pi
^{ij}$, i.e., the metric $F(r)$ and angular momentum $p(r)$.

In order to do that, one integrates the Hamiltonian constraint ${\cal H}%
_{\bot }$ across the shell in a constant-time slice $\Sigma _{t}$. By means
of suitable projections between a timelike (ADM) foliation and the commoving
frame $\{\lambda ,X\}$ on the hypersurface $\Sigma _{\xi }$ generated by the
shell, one is able to obtain a relation between the radial velocity and the
difference in the function of the metric $\triangle F^{2}$. For the matter
stress tensor $T_{\bot \bot }$ in Eq.(\ref{hper}) we take a perfect fluid (%
\ref{Tmunu}) with an energy density $\sigma $ and a shell tension $\tau $.
Then, the equation of motion for the shell is given by

\begin{equation}
-\triangle F^{2}=\pi R(\alpha _{+}+\alpha _{-})\left\{ \gamma ^{2}\sigma
-\tau (\gamma ^{2}-1)\right\}  \label{deltaF2}
\end{equation}%
with $\alpha $ defined in Eq.(\ref{alphadef}).

The integration of the Hamiltonian constraint (evaluated in a minisuperspace
model) across the shell is performed first for a finite thickness $%
\varepsilon $. In this procedure, the l.h.s. of Eq.(\ref{deltaF2}) comes as
the only nonvanishing term after the thin shell limit $\varepsilon
\rightarrow \infty $ is taken, because all the rest have finite
discontinuities across the shell. The same argument can be applied to the
electromagnetic part of the stress tensor, that does not contribute to the
integration of $T_{\bot \bot }$ and therefore, the above equation remains
the same when the shell also carries electric charge.

In addition, the key point to integrate the angular component of the
Hamiltonian ${\cal H}_{\varphi }$ is the adoption of the matching condition $%
N^{\varphi }\left( R\right) =0$. In simple terms, this condition means that
the angular shift is chosen as zero over the shell at any time, being the
only choice that permits a commoving observer (rotating and falling with the
shell) to have continuity in the angular displacement across the shell.
Performing this integration leads the difference in angular momentum between
the interior and exterior regions

\begin{equation}
-2\Delta p=\triangle J=2\pi \gamma ^{2}R^{3}\Omega (\sigma -\tau ).
\label{deltaJ}
\end{equation}%
As in the radial collapse case, it is useful to express Eq. (\ref{deltaF2})
in the form

\begin{equation}
\alpha _{+}-\alpha _{-}=-\pi R\left\{ \gamma ^{2}\sigma -\tau (\gamma
^{2}-1)\right\}  \label{deltalpha}
\end{equation}%
that has been obtained from multiplying (\ref{deltaF2}) by $\alpha
_{+}-\alpha _{-}$. This relation can be better understood in the nonrotating
limit ($\gamma =1$), where it matches the standard Israel formula for radial
collapse dynamics that comes from the discontinuity of the radial
acceleration across the shell.

The set of equations (\ref{deltalpha}) and (\ref{deltaJ}) is not enough to
completely determine the dynamic evolution of the shell, as we required an
equation of state for the shell matter.

For simplicity, we are going to illustrate the method here with the collapse
of a thin shell made of coherent dust. Thus, we take the tension $\tau =0$
in the Eqs. (\ref{deltalpha}) and (\ref{deltaJ})

\begin{eqnarray}
\alpha _{+}-\alpha _{-} &=&\sqrt{F_{+}^{2}(R)+\dot{R}^{2}}-\sqrt{%
F_{-}^{2}(R)+\dot{R}^{2}}=-\frac{m\gamma }{2}  \label{disalpha2} \\
J &=&m\gamma \Omega R^{2}  \label{Jext}
\end{eqnarray}%
where $m=2\pi \gamma R\sigma $ is a conserved quantity in the commoving
frame (rest-frame mass) and the momentum $p_{+}=-J/2$ and the factor $\gamma
$ takes the explicit form
\begin{equation}
\gamma (R)=\left( 1+\frac{J^{2}}{m^{2}R^{2}}\right) ^{-1/2}.  \label{gammaJ}
\end{equation}

In most of cases of thin shell collapse it is convenient to get $\dot{R}%
^{2}(\lambda )$ from the expressions (\ref{disalpha2}) and (\ref{Jext}) by
quadrature, in order to carry out an {\em effective potential }analysis and
to conclude in which regions of the space the shell motion is possible. In
this picture, it is clear that the shell parameters $m$ and $q$ and the set
of initial conditions ($\dot{R}_{0},$ $R_{0},$ $\Omega _{0}$) should
completely determine the parameters $M$, $J$ and $Q$ of the outer black hole
and the further shell collapse.

However, for the present set-up, both functions $F_{-}$ and $F_{+}$ --given
by Eqs. (\ref{Fminus}) and (\ref{Fplus}), respectively-- are defined only
implicitly in terms of different radial coordinates $\rho $ and $\tilde{r}$.
This makes difficult to obtain an analytical formula for the velocity
because it requires to invert the formulas (\ref{r_rho}) and (\ref{r_rtilde}%
) to get the relations $\rho (r)$ and $\tilde{r}(r)$.

For a simple choice of the initial conditions, we may relate some of the
above parameters. For instance, if the shell is released with radial
velocity $\dot{R}_{0}$ equal to zero, it is possible to express the constant
$\Theta $ present in the magnetic field in terms of the initial position $%
R_{0}$ and angular velocity $\Omega _{0}$ where, combining Eqs. (\ref%
{Bminusq}) and (\ref{Bminuspart})
\begin{equation}
\Theta =\frac{q\Omega _{0}R_{0}}{2\pi }
\end{equation}%
and the angular momentum $J$ of the outer solution is given by Eq(\ref%
{deltaJ}) as%
\begin{equation}
J=\frac{R_{0}^{2}\Omega _{0}}{\sqrt{1-R_{0}^{2}\Omega _{0}^{2}}}m.
\end{equation}
Nevertheless, the mass $M$ of the outer solution is difficult to obtain from
direct substitution of the initial conditions in Eq.(\ref{disalpha2}). It is
true that, in principle, we can get an expression for $M(\tilde{M},\tilde{Q}%
,\omega )$ from the set of Eqs.(\ref{Mfromtildes}-\ref{Qfromtildes}), but
the boost parameter $\omega $ also depends on the exterior parameters
through Eq.(\ref{omegacubic}).

As an alternative to the effective potential method, we can differentiate
Eq. (\ref{disalpha2}) respect the time coordinate $\lambda $. Thus, we find
that the radial acceleration satisfies

\begin{equation}
m\gamma \frac{d^{2}R}{d\lambda ^{2}}=\alpha _{+}\left( F_{-}^{2}\right)
^{\prime }-\alpha _{-}\left( F_{+}^{2}\right) ^{\prime }-m\alpha _{+}\alpha
_{-}\frac{d\gamma }{dR}
\end{equation}%
where the prime denotes the radial derivative.

Infinitesimally close to the shell position $r=R(\lambda )$, at either side
of it, the set of equations (\ref{Fprime})-(\ref{varAphi}) are still valid
for the corresponding region. In particular, Eq.(\ref{Fprime}) relates the
derivative of the function $F^{2}$ in the metric with the remaining fields
and parameters

\begin{eqnarray}
\left( F_{-}^{2}\right) ^{\prime } &=&2R-\frac{B_{-}^{2}R}{2} \\
\left( F_{+}^{2}\right) ^{\prime } &=&2R-\frac{J^{2}}{2R^{3}}-\frac{q^{2}}{2R%
}
\end{eqnarray}%
where the outer total charge is the one carried by the shell $Q_{+}=q$, and
the magnetic field vanishes just outside the shell. After some
rearrangements, the acceleration can be written as

\begin{equation}
m\gamma \frac{d^{2}R}{d\lambda ^{2}}=-m\gamma R-\alpha _{+}\frac{B_{-}^{2}R}{%
2}+\alpha _{-}\frac{q^{2}}{2R}+\alpha _{-}^{2}\frac{J^{2}}{m\gamma R^{3}}
\label{motion}
\end{equation}%
using the formula (\ref{disalpha2}) in a convenient manner.

What we finally obtain is a nonlinear differential equation (\ref{motion})
that governs the shell collapse. Even though there is no an analytical
expression for $R$ as a function of time, the equation should provide the
shell dynamics and can be analyzed numerically for different values of $M$, $%
J$ and $Q$ of the external black hole. This is out of the scope of this
paper, where we have focused on analytical aspects of this gravitational
collapse process. We hope to report this issue elsewhere.

\section{Conclusions}

In this paper we have developed an application of a Hamiltonian treatment
for the collapse of rotating thin shells in three dimensions. MTZ black hole
solutions are obtained as the result of the collapse of a charged spinning
ring.

As the shell crosses its own event horizon, it leaves behind a vanishing
magnetic field at that position, a desirable feature when we are interested
in the creation of charged rotating black holes in the outer region.

We have dealt so far with only one possible mechanism to create an external
MTZ black hole. However, we can also consider other processes, for instance,
the consecutive collapse of two shells: one carrying the electric charge and
another the angular momentum (or viceversa) and see the physical
implications. Generating a charged rotating solution in this form may have
the advantage of a better understanding of the internal structure of the
black hole. For instance, we could study the stability of the Cauchy horizon
to the addition of another {\em hair} ($Q$ or $J$) \cite{ori, husain},
working in Kruskal-like coordinates, what cannot be done in the present
paper where the formalism is valid only until $r=r_{+}$ (Schwarzschild-like
coordinates).

Finally, one might naturally wonder whether it is possible to generate this
collapse process by the action of the rotational boost discussed in Refs.%
\cite{MTZ, Clement96}, starting from the radial, electrically charged
collapse. The answer to this question is negative because a boost acting on
the interior and exterior patches with the same parameter $\omega $ cannot
produce arbitrary solutions parameters. In particular, this procedure does
not reproduce the case presented in this paper, because it necessarily
induces angular momentum in the interior zone. At best, even if one could
redefine the radial coordinate in the interior and exterior regions to
obtain rotating solutions, the intrinsic metric of the shell under the
action of the boost does not preserve its form and thus, it is not clear how
to impose the matching conditions on the shell position.

\section{Acknowledgments}

I wish to thank Juan Cris\'{o}stomo and Claudio Teitelboim for enlightening
comments and helpful discussions. I also thank M\'{a}ximo Ba\~{n}ados and
Jorge Zanelli for useful conversations. Institutional support to Centro de
Estudios Cient\'{\i}ficos (CECS) from Empresas CMPC is acknowledged. CECS is
a Millennium Science Institute and is funded in part by grants from Fundaci%
\'{o}n Andes and the Tinker Foundation. This work was partially funded by
the grant 3030029 from FONDECYT.

\end{document}